\title{Recursive List Decoding for Reed-Muller Codes\;\\
and Their Subcodes}

\author{Ilya Dumer and Kirill Shabunov\thanks{This research was
supported by the NSF grant CCR-0097125.}\\
College of Engineering\\
University of California, Riverside\\
Riverside, CA 92521\\
{\tt dumer@ee.ucr.edu}\\
{\tt shabunov@ee.ucr.edu}
}
\date{January 30, 2002}

\documentclass[12pt]{article}
\usepackage{latexsym}
\usepackage{amsmath}
\usepackage{graphicx}
\usepackage{amsfonts}
\usepackage{amssymb}

 \newtheorem{theo}{Theorem}[section]
 \newtheorem{lemma}{Lemma}[section]

\begin{document}
\parindent=10pt
\maketitle

\begin{abstract}
We consider recursive decoding for Reed-Muller (RM) codes and their subcodes.
Two new recursive techniques are described. We analyze asymptotic properties
of these algorithms and show that they substantially outperform other decoding
algorithms with nonexponential complexity known for RM codes. Decoding
performance is further enhanced by using intermediate code lists and
permutation procedures. For moderate lengths up to 512, near-optimum decoding
with feasible complexity is obtained.
\end{abstract}


\section{Introduction}

Below we consider Reed-Muller (RM) codes and their
subcodes. We use notation $\left\{\genfrac{}{}{0pt}{}{m}{r}\right\}$
for RM codes of length $n=2^{m},$ dimension $k=\sum_{i=0}%
^{r}{\binom{m}{i}}$ and distance $d=2^{m-r}$. In this paper, we wish to design
fast decoding algorithms that outperform other algorithms known for RM codes.
To achieve this goal, we will use \textit{recursive techniques}
that make decoding decisions by combining decoding results obtained on the
shorter codes.

RM codes are almost on par with the best codes on moderate lengths $n\leq128$
and have found numerous applications thanks to fast decoding procedures.
First, \textit{majority algorithm }was developed in \cite{ree} followed later
by numerous developments. Such a decoding has low complexity of the order $nk$
and enables bounded distance decoding.

It is also important that majority decoding substantially extends
bounded-distance \textit{threshold} of $d/2.$ Given  an infinite sequence of
codes $A_{i}(n_{i},d_{i}),$ we say that a decoding algorithm $\Psi$ achieves
decoding thresholds $\delta_{i}$ if for any $\epsilon>0$ only a vanishing
fraction of error patterns of weight $\delta_{i}(1-\epsilon)$ is left
uncorrected as $n_{i}\rightarrow\infty.$ It can be proven \cite{kri} that
majority algorithm achieves a threshold
\begin{equation}
\delta=(d\ln d)/4\ \label{delta-maj}%
\end{equation}
for long RM codes of fixed rate $R$ (here and below we omit index $i$).
For long low-rate RM codes of fixed order $r$, it is
shown \cite{dum3} that majority decoding achieves a threshold
\[
\delta=n(1-\varepsilon_{r}^{\text{maj}})/2,
\]
where the residual term $\varepsilon_{r}^{\text{maj}}$ has a slowly declining
order
\begin{equation}
\varepsilon_{r}^{\text{maj}}\sim(m/d)^{1/2^{r+1}}=(m/2^{m-r})^{1/2^{r+1}%
},\quad m\rightarrow\infty.\label{eps-maj}%
\end{equation}
Therefore $\delta$ exceeds bounded distance threshold $d/2$ approximately
$2^{r}$ times and approaches the upper limit of $n/2$ for long codes.

Another efficient technique is based on recursive algorithms of \cite{lit} and
\cite{kab}. The recursive structure of RM codes
$\left\{\genfrac{}{}{0pt}{}{m}{r}\right\}$
is well known \cite{MS} and is formed by the\textit{ Plotkin
construction} $(\mathbf{u,u+v}),$ where subblocks $\mathbf{u}$ and
$\mathbf{v}$ are taken from the codes
$\left\{\genfrac{}{}{0pt}{}{m-1}{r}\right\}$
and $\left\{\genfrac{}{}{0pt}{}{m-1}{r-1}\right\}$.
It is proven in \cite{lit} and \cite{kab} that this recursive
structure allows to execute bounded distance decoding with the lowest
complexity order of $n\min(r,m-r)$ known for RM codes. The techniques
developed below also show that recursive algorithms of \cite{lit} and
\cite{kab} are on par with majority decoding and achieve the same
error-correcting thresholds $\delta$ in both cases (\ref{delta-maj}) and
(\ref{eps-maj}).

One more efficient algorithm based on permutation decoding was designed in
\cite{sid} for RM codes $\left\{\genfrac{}{}{0pt}{}{m}{2}\right\}$.
This algorithm gives a slightly higher complexity order of
$n^{2}m$ while reducing the corresponding residual term $\varepsilon
_{2}^{\text{maj}}$ from (\ref{eps-maj}) to the lower order of $(m/n)^{1/4}$
as $m\rightarrow\infty.$

The above algorithms can also be extended for soft decision channels. For RM
codes of fixed rate $R,$ soft decision majority decoding \cite{dum3} gives a
threshold of Euclidean weight $\rho=(n/m)^{1/2^{r+1}}\sqrt{n}$. Using
technique of \cite{dum3}, it can be proven that recursive algorithms of
\cite{lit} and \cite{kab} also have the same error-correcting threshold
$\rho.$ For RM codes $\left\{\genfrac{}{}{0pt}{}{m}{2}\right\}$,
the algorithm of \cite{sid} allows to increase the Euclidean
threshold $\rho$ to the order of $(n/m)^{1/4}\sqrt{n}.$ Finally, multistage
maximum-likelihood decoding is performed in \cite{for} by designing an
efficient multilevel trellis structure. ML decoding supersedes recursive
algorithms. In particular, it can be proven (similarly to \cite{sid}) that ML
decoding has Euclidean threshold upper bounded by the order of $n\sqrt
{(r!\ln2)/2m^{r}}.$ However, ML decoding complexity is exponential in $n$.

Below we design new recursive algorithms that substantially outperform other
(nonexponential) algorithms known for RM codes for both hard and soft decision
channels. Our basic recursive procedure will split the RM code
$\left\{\genfrac{}{}{0pt}{}{m}{r}\right\}$
of length $n$ into two RM codes of length $n/2$. Decoding is then
relegated further to the shorter codes until we reach basic codes and perform
ML decoding with complexity $O(n\log n).$ In all intermediate steps, we only
recalculate the reliabilities of the newly defined symbols.

Below in Section 2 we consider recursive structure of RM codes in more detail.
Then in Sections 3 and 4 we proceed with decoding techniques and design two
different versions $\Psi_{r}^{m}$ and $\Phi_{r}^{m}$ of our recursive
algorithm. In Section 5 we proceed with further improvements. In particular,
decoding performance will be considerably improved by using subcodes of RM
codes. Another improvement is based on using relatively short lists of
codewords in the intermediate steps of the recursion. Finally, we use
different permutations taken from the symmetry (automorphism) group of the
code. As a result, we closely approach ML decoding performance on the
blocklength of 256 and for low-rate codes of length 512.

\section{Recursive structure}

Recursive techniques known for RM codes are based on the \textit{Plotkin
construction}. Here any RM code $\left\{
\genfrac{}{}{0pt}{}{m}{r}%
\right\}  $ is represented in the form $(\mathbf{u,u+v}),$ where $\mathbf{u}$
and $\mathbf{v}$ are two subblocks of length $2^{m-1}$ that run
through the codes $\left\{
\genfrac{}{}{0pt}{}{m-1}{r}%
\right\}  $ and $\left\{
\genfrac{}{}{0pt}{}{m-1}{r-1}%
\right\}  ,$ respectively. By continuing this process on codes $\left\{
\genfrac{}{}{0pt}{}{m-1}{r}%
\right\}  $ and $\left\{
\genfrac{}{}{0pt}{}{m-1}{r-1}%
\right\}  ,$ we obtain RM codes of length $2^{m-2}$ and so on. Finally, we
arrive at the end nodes, which are repetition codes $\left\{
\genfrac{}{}{0pt}{}{g}{0}%
\right\}  $ for any $g=1,\ldots,m-r$ and full spaces $\left\{
\genfrac{}{}{0pt}{}{h}{h}%
\right\}  $ for any $h=1,\ldots,r.$ This is schematically shown in Fig. 1 for RM
codes of length 16. In Fig.~2, we consider incomplete decomposition for codes
of length 32 terminated at the biorthogonal codes and single-parity check
codes.
\smallskip

\ \ \ \ \ \ \ \ \ \ \ \ \ \ \ \ \ \ \ \ \ \ \ \ \ \ \ \ \ \ \ {\small 0,0}%
\ \ \ \ \ \ \ \ \ \ \ \ \ \ \ \ \ \ \ \ \ \ \ \ \ \ \ \ \ \ \ \ \ \ \ \ \ \ \ \ \ \ \ \ \ \ \ \ \ \ \ \ \ {\small 2,1}%
\

\ \ \ \ \ \ \ \ \ \ \ \ \ \ \ \ \ \ \ \ \ \ \ \ \ \ \ \ $\nearrow$
$\ \nwarrow$%
\ \ \ \ \ \ \ \ \ \ \ \ \ \ \ \ \ \ \ \ \ \ \ \ \ \ \ \ \ \ \ \ \ \ \ \ \ \ \ \ \ \ \ \ \ \ \ \ \ $\nearrow
$ $\ \nwarrow$

\ \ \ \ \ \ \ \ \ \ \ \ \ \ \ \ \ \ \ \ \ \ \ \ \ {\small 1,0}{\large \ \ \ }%
\ {\large \ }\ {\small 1,1 }%
\ \ \ \ \ \ \ \ \ \ \ \ \ \ \ \ \ \ \ \ \ \ \ \ \ \ \ \ \ \ \ \ \ \ \ \ \ \ \ \ \ \ {\small 3,1}%
{\large \ \ \ }\ {\large \ }\ \ {\small 3,2}

$\ \ \ \ \ \ \ \ \ \ \ \ \ \ \ \ \ \ \ \ \ \ \nearrow$ \ $\nwarrow
${\large \ \ \ }\ $\nearrow$ $\ \nwarrow$
$\ \ \ \ \ \ \ \ \ \ \ \ \ \ \ \ \ \ \ \ \ \ \ \ \ \ \ \ \ \ \ \ \ \ \ \ \nearrow
$ \ $\nwarrow${\large \ \ \ }\ $\nearrow$ $\ \nwarrow$

\ \ \ \ \ \ \ \ \ \ \ \ \ \ \ \ \ \ {\small 2,0}{\large \ \ \ }\ {\large \ \ }%
\ {\small 2,1}{\large \ \ \ }\ {\large \ \ }\ \ {\small 2,2
\ \ \ \ \ \ \ \ \ \ \ \ \ \ \ \ \ \ \ \ \ \ \ \ \ \ \ \ \ \ \ 4,1{\large \ \ \ }%
\ {\large \ \ \ \ }4,2{\large \ \ \ }\ {\large \ \ }\ 4,3\ \ \ \ \ \ \ \ }

\ \ \ \ \ \ \ \ \ \ \ \ \ \ \ \ $\nearrow$ $\ \nwarrow${\large \ \ \ }%
\ $\nearrow$ $\ \nwarrow${\large \ \ \ }\ $\nearrow$ $\ \nwarrow
$\ \ \ \ \ \ \ \ \ \ \ \ \ \ \ \ \ \ \ \ \ \ \ \ $\nearrow$ $\ \nwarrow
${\large \ \ \ }\ $\nearrow$ $\ \nwarrow${\large \ \ \ }\ $\nearrow$
$\ \nwarrow$ \

\ \ \ \ \ \ \ \ \ \ \ \ \ {\small 3,0}{\large \ \ \ \ \ \ }{\small 3,1}%
{\large \ \ \ \ \ \ \ }\ {\small 3,2}{\large \ \ \ }\ {\large \ \ }%
\ {\small 3,3 \ \ \ \ \ \ \ \ \ \ \ \ \ \ \ \ \ \ \ \ \ 5,1{\large \ \ \ }%
\ {\large \ \ }5,2{\large \ \ \ \ \ \ }\ \ 5,3{\large \ \ \ \ \ \ }5,4\ }

\ \ \ \ \ \ \ \ \ $\nearrow$ $\ \nwarrow${\large \ \ \ }\ $\nearrow$
$\ \nwarrow${\large \ \ \ }\ $\nearrow$\ $\ \nwarrow${\large \ \ \ }%
\ $\nearrow$ $\ \nwarrow$\ \ \ \ \

\ \ \ \ \ \ {\small 4,0}{\large \ \ \ \ \ \ }\ {\small 4,1}{\large \ \ \ }%
\ {\large \ }\ \ \ {\small 4,2}{\large \ \ \ }\ {\large \ \ }\ \ {\small 4,3}%
{\large \ \ \ }\ {\large \ \ }\ {\small 4,4 \ \ \ \ \ \ }

\vspace{0.1in}

\quad Fig. 1$:$ Full decomposition\qquad\qquad\qquad\qquad\qquad Fig. 2$:$
Partial decomposition

\vspace{0.1in}
\setcounter{figure}{2}
This recursive structure is also exhibited in the generator matrices of RM codes.
As an example, on Fig.~\ref{ds_figG0307} we present a generator matrix for
$\left\{\genfrac{}{}{0pt}{}{7}{3}\right\}$ code.
\begin{figure}[htb]
\begin{center}
\includegraphics[width=\linewidth]{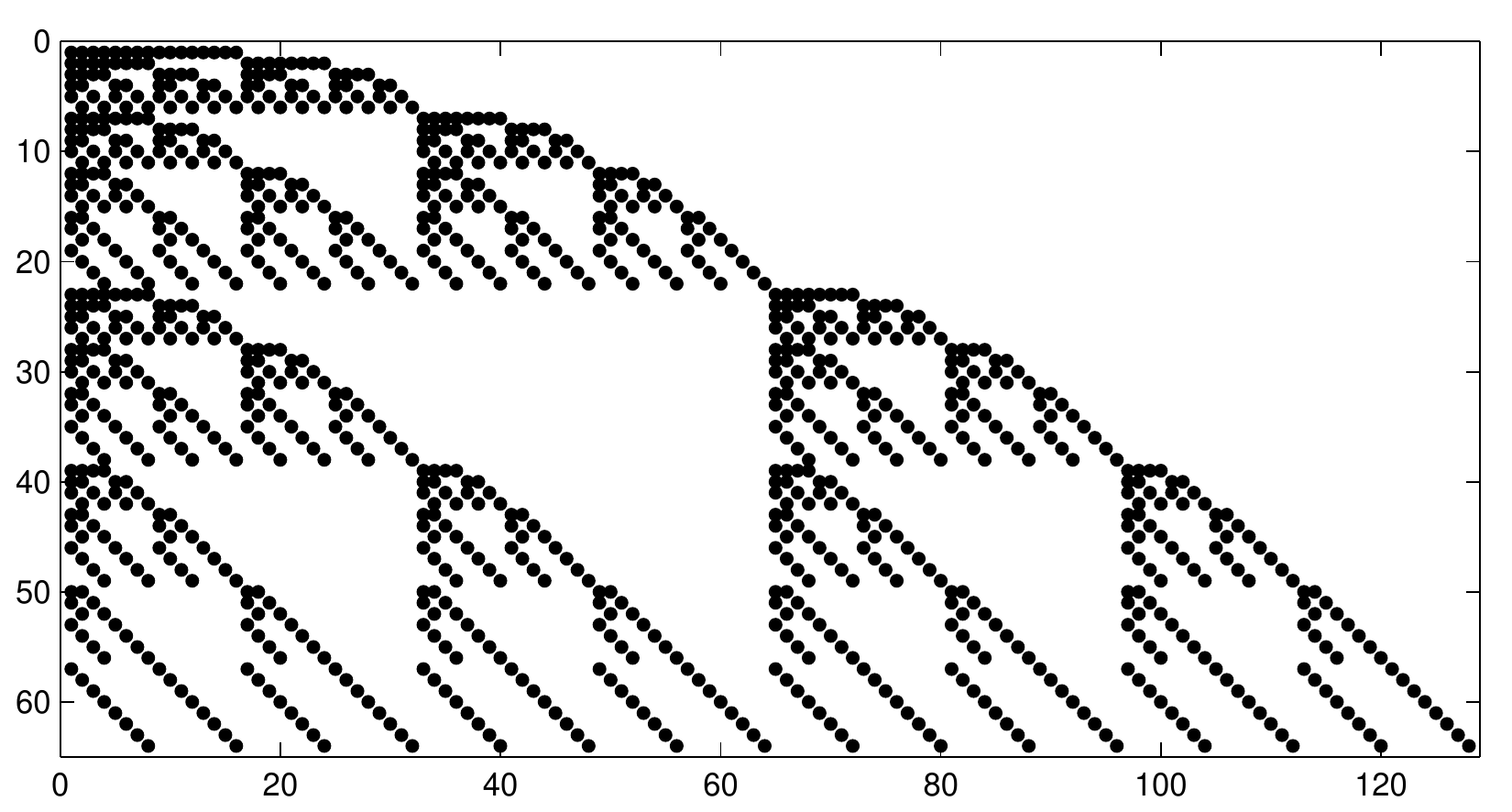}
\caption{A generator matrix of $\left\{\genfrac{}{}{0pt}{}{7}{3}\right\}$ code.
Ones are denoted by dots, zeroes are not shown.}
\label{ds_figG0307}
\end{center}
\end{figure}

Now let $\mathbf{a}\genfrac{}{}{0pt}{}{m}{r}=\{a_{j}|j=1,k\}$
be a block of information bits $a_{j}$ that encodes a
vector $(\mathbf{u},\mathbf{u}+\mathbf{v}).$ Then recursive procedure splits
$\mathbf{a}\genfrac{}{}{0pt}{}{m}{r}$ into two information subblocks
$\mathbf{a}\genfrac{}{}{0pt}{}{m-1}{r}$ and $\mathbf{a}\genfrac{}{}{0pt}{}{m-1}{r-1}$
that encode vectors $\mathbf{u}$ and $\mathbf{v,}$ respectively. In this
way, information subblocks are split until we arrive at the end nodes$.$ Thus,
any specific codeword can be encoded from the information strings assigned to
the end nodes $\left\{\genfrac{}{}{0pt}{}{g}{0}\right\}$ or
$\left\{\genfrac{}{}{0pt}{}{h}{h}\right\}$.

Also, it can be proven that recursive encoding of code
$\left\{\genfrac{}{}{0pt}{}{m}{r}\right\}$ has complexity
\begin{equation}
\psi_{r}^{m}\leq n\min(r,m-r). \label{encoding}%
\end{equation}
This observation comes from the two facts. First, the end nodes
$\left\{\genfrac{}{}{0pt}{}{g}{0}\right\}$ and
$\left\{\genfrac{}{}{0pt}{}{h}{h}\right\}$
satisfy the complexity bound (\ref{encoding}). Second, we can
obtain an $(\mathbf{u,u+v})$-codeword using two constituent codewords taken
from $\left\{\genfrac{}{}{0pt}{}{m-1}{r-1}\right\}$
and $\left\{\genfrac{}{}{0pt}{}{m-1}{r}\right\}$.
Therefore the overall complexity satisfies inequality
$\psi_{r}^{m}\leq\psi\genfrac{}{}{0pt}{}{m-1}{r-1} +
\psi\genfrac{}{}{0pt}{}{m-1}{r}+\frac{n}{2}$.
Now we see that code $\left\{\genfrac{}{}{0pt}{}{m}{r}\right\}$
satisfies (\ref{encoding}) if the two constituent codes
do. In particular, for $r<m/2,$ we obtain the bound
\[
\psi_{r}^{m}\leq n(r-1)/2+nr/2+n/2=nr.
\]

Now consider an information bit $a_{j}$ associated with any left node, say
$\left\{
\genfrac{}{}{0pt}{}{g}{0}%
\right\}  .$ Then splitting procedure also allows us to map this symbol
$a_{j}$ onto a specific ``binary path'' $\overline{j}=(j_{1},\ldots,j_{m})$
leading from the origin $\left\{
\genfrac{}{}{0pt}{}{m}{r}%
\right\}  $ to the end node $\left\{
\genfrac{}{}{0pt}{}{g}{0}%
\right\}  $. To do this, we first take $j_{1}=0$ for any $a_{j}\in$
$\mathbf{a}%
\genfrac{}{}{0pt}{}{m-1}{r}%
$ and $j_{1}=1$ for $\mathbf{a}%
\genfrac{}{}{0pt}{}{m-1}{r-1}%
.$ On any step $s=2,\ldots,m$, we take $j_{s+1}=1$ when moving to the left (say,
from $\mathbf{a}%
\genfrac{}{}{0pt}{}{m-1}{r-1}%
$ to $\mathbf{a}%
\genfrac{}{}{0pt}{}{m-2}{r-2}%
$\textbf{ }$)$ or $j_{s+1}=0$ when moving to the right. Then we get a subindex
$=(j_{1},\ldots,j_{m-g})$ at the left-end node $\left\{
\genfrac{}{}{0pt}{}{g}{0}%
\right\}  $ associated with the information bit $a_{j}$. We complete this
mapping by adding $g$ zeros $j_{m-g+1}=\cdots=j_{m}=0$ to $\overline{j}_{g}.$ As
a result, we obtain the full path $\overline{j}$ that arrives at the node
$\left\{
\genfrac{}{}{0pt}{}{0}{0}%
\right\}  .$

Now consider the right-end node $\left\{
\genfrac{}{}{0pt}{}{h}{h}%
\right\}  $ that includes $2^{h}$ information bits. In this case, the same
mapping procedure gives a subpath $\overline{j}_{h}=(j_{1},\ldots,j_{m-h})$. To
enumerate any specific information bit $a_{j}$ associated with this node
$\left\{
\genfrac{}{}{0pt}{}{h}{h}%
\right\}  $, subindex $\overline{j}_{h}=(j_{1},\ldots,j_{m-h})$ is appended by
any combination $(j_{m-h+1},\ldots,j_{m})$. As a result, we enumerate all $2^{h}$
bits $a_{j}$ given at the node $\left\{
\genfrac{}{}{0pt}{}{h}{h}%
\right\}  .$ It can also be seen that all indices $\{\overline{j}\}$ include
at most $r$ ones in their binary representation. Therefore all $k$ information
bits of the whole code are mapped onto $m$-digital binary paths $\overline{j}$
of weight $r$ or less.

\section{New decoding techniques}

\noindent Below we consider a new recursive algorithm $\Psi_{\text{rec}}$
based on the $(\mathbf{u,u+v})$ construction. The received block
$\mathbf{y}=(\widetilde{\mathbf{u}},\widetilde{\mathbf{u+v}})$ consists of two
halves $\mathbf{y}^{\prime}$ and $\mathbf{y}^{\prime\prime}$ corrupted by
noise. By taking outputs $y_{i}^{\prime}$ and $y_{i}^{\prime\prime}$, the
decoder finds the posterior probabilities of symbols $u_{i}$ and
$u_{i}\mathbf{+}v_{i}:$
\[
p_{i}^{\prime}\overset{\text{def}}{=}\Pr\{u_{i}=0\,\,|\,\,y_{i}^{\prime
}\},\quad p_{i}^{\prime\prime}\overset{\text{def}}{=}\Pr\{u_{i}\mathbf{+}%
v_{i}=0\,\,|\,\,y_{i}^{\prime\prime}\},\quad i=1,\ldots,n/2.
\]
We first try to find the better protected codeword $\mathbf{v}$ from
$\left\{
\genfrac{}{}{0pt}{}{m-1}{r-1}%
\right\}  \mathbf{.}$ Then we decode the block $\mathbf{u\in}\left\{
\genfrac{}{}{0pt}{}{m-1}{r}%
\right\}  $.

\textbf{Step 1}. To find a subblock $\mathbf{v}$ in hard-decision decoding,
one would use its corrupted version $\widetilde{\mathbf{v}}=\widetilde
{\mathbf{u}}+\widetilde{\mathbf{u}+\mathbf{v}}.$ Using more general approach,
we find $n/2$ posterior probabilities
\begin{equation}
p_{i}^{\mathbf{v}}\overset{\text{def}}{=}\Pr\{v_{i}=0\,\,|\,\,y_{i}^{\prime
}\,,y_{i}^{\prime\prime}\}=p_{i}^{\prime}p_{i}^{\prime\prime}+(1-p_{i}%
^{\prime})(1-p_{i}^{\prime\prime}). \label{1}%
\end{equation}
Here we apply the formula of total probability to the binary sum of
independent symbols $u_{i}$ and $u_{i}\mathbf{+}v_{i}.$ Now we can use any
soft-decision decoding $\Psi_{\mathbf{v}}(p_{i}^{\mathbf{v}})$ to find vector
$\widehat{\mathbf{v}}$. This completes Step 1 of our algorithm. Vector
$\widehat{\mathbf{v}}$ is then passed to Step 2.

\textbf{Step 2}. Now we use both symbols $y_{i}^{\prime\prime}$ and $v_{i}$
to estimate symbol $u_{i}$ on the right half. \textit{Assuming that}
$\widehat{\mathbf{v}}=\mathbf{v}$ , we find that symbol $u_{i}$ has
posterior probability
\[
p_{i}^{\wedge}\overset{\text{def}}{=}\Pr\{u_{i}=0\,\,|\,y_{i}^{\prime\prime
},\,\widehat{v}_{i}\}=\left\{
\begin{array}
[c]{ll}%
p_{i}^{\prime\prime}, & \text{if}\;\widehat{v}_{i}=0,\\
1-p_{i}^{\prime\prime}, & \text{if}\;\widehat{v}_{i}=1.
\end{array}
\right.
\]
Now we have the two posterior probabilities $p_{i}^{\prime}$ and
$p_{i}^{\wedge}$ of symbol $u_{i}$ obtained on both corrupted halves. By using
the Bayes' formula, we find the combined estimate
\begin{equation}
p_{i}^{\mathbf{u}}\overset{\text{def}}{=}\Pr\{u_{i}=0\,|\,\,p_{i}^{\prime
},\,p_{i}^{\wedge}\}=\frac{p_{i}^{\prime}p_{i}^{\wedge}}{p_{i}^{\prime}%
p_{i}^{\wedge}+(1-p_{i}^{\prime})(1-p_{i}^{\wedge})}. \label{3}%
\end{equation}
Finally, we perform soft decision decoding $\Psi_{\mathbf{u}}(p_{i}%
^{\mathbf{u}})$ and find a subblock $\widehat{\mathbf{u}}\mathbf{\in
}\left\{
\genfrac{}{}{0pt}{}{m-1}{r}%
\right\}  .$ So, the basic soft-decision decoding $\Psi_{\text{rec}}$ uses
procedures~$\Psi_{\mathbf{v}}$,~$\Psi_{\mathbf{u}}$ and outputs a decoded
codeword~$\mathbf{\hat{c}}$ and the corresponding information block~$\widehat
{\mathbf{a}}$ as follows.
\[
\frame{$%
\begin{array}
[c]{l}%
\text{Algorithm }\Psi_{\text{rec}}.\\
\text{1. Calculate probabilities }p_{i}^{\mathbf{v}}\text{ according to
(\ref{1}).}\\
\text{2. Decode }\mathbf{\hat{v}}\text{ using }\Psi_{\mathbf{v}}%
(p_{i}^{\mathbf{v}}),i=1,\ldots,n/2.\\
\text{3. Calculate probabilities }p_{i}^{\mathbf{v}}\text{ according to
(\ref{3}).}\\
\text{4. Decode }\mathbf{\hat{u}}\text{ using }\Psi_{\mathbf{u}}%
(p_{i}^{\mathbf{u}}),i=1,\ldots,n/2.\\
\text{5. Output decoded components:}\\
\qquad\widehat{\mathbf{a}}:=(\widehat{\mathbf{a}}_{\mathbf{v}}\mid
\widehat{\mathbf{a}}_{\mathbf{u}});\quad\mathbf{\hat{c}}:=(\mathbf{\hat{u}%
}\mid\mathbf{\hat{u}}+\mathbf{\hat{v}})\qquad
\end{array}
$}%
\]

\noindent In a more general scheme $\Psi_{r}^{m}$, we repeat this recursion by
decomposing subblocks $\widehat{\mathbf{v}}$ and $\widehat{\mathbf{u}}$
further. On all intermediate steps, we only recalculate the probabilities of
the newly defined symbols. Finally, we perform ML decoding once we reach the
end nodes $\left\{
\genfrac{}{}{0pt}{}{g}{0}%
\right\}  $ and $\left\{
\genfrac{}{}{0pt}{}{h}{h}%
\right\}  $. The algorithm is described below.
\[
\frame{$%
\begin{array}
[c]{l}%
\text{Algorithm }\Psi_{r}^{m}.\\
\text{1. If }0<r<m\text{, find }\Psi_{\text{rec}}(\mathbf{p)}\text{ using}\\
\Psi_{\mathbf{v}}=\Psi_{r-1}^{m-1}\text{ and }\Psi_{\mathbf{u}}=\Psi_{r}%
^{m-1}\text{.}\\
\text{2. If }r=0\text{ decode code }\left\{
\genfrac{}{}{0pt}{}{r}{0}%
\right\}  .\\
\text{3. If }r=m\text{ decode code }\left\{
\genfrac{}{}{0pt}{}{r}{r}%
\right\}  .
\end{array}
$}%
\]

\noindent In the next algorithm $\Phi_{r}^{m}$, we terminate decoding
$\Psi_{\mathbf{v}}$ at the biorthogonal codes $\left\{
\genfrac{}{}{0pt}{}{g}{1}%
\right\}  $.
\[
\frame{$%
\begin{array}
[c]{l}%
\text{Algorithm }\Phi_{r}^{m}.\\
\text{1. If }1<r<m\text{, find }\Psi_{\text{rec}}(\mathbf{p)}\text{ using}\\
\Psi_{\mathbf{v}}=\Psi_{r-1}^{m-1}\text{ and }\Psi_{\mathbf{u}}=\Psi_{r}%
^{m-1}\text{.}\\
\text{2. If }r=1\text{ decode code }\left\{
\genfrac{}{}{0pt}{}{r}{1}%
\right\}  .\\
\text{3. If }r=m\text{ decode code }\left\{
\genfrac{}{}{0pt}{}{r}{r}%
\right\}  .
\end{array}
$}%
\]

Thus, procedures $\Psi_{r}^{m}$ and $\Phi_{r}^{m}$ have recursive structure
that calls itself until ML decoding is applied on the end nodes. ML decoding
of biorthogonal codes has complexity order of $n\log_{2}n$. Simple analysis
also shows that recalculating all posterior probabilities in (\ref{1}) and
(\ref{3}) has complexity at most $5n.$ Therefore our decoding complexity
$\Phi_{r}^{m}$ satisfies recursion
\[
\Phi_{r}^{m}\leq\Phi_{r-1}^{m-1}+\Phi_{r}^{m-1}+5n.
\]
Similarly to the derivation of (\ref{encoding}), this recursion brings the
overall complexity of $\Phi_{r}^{m}$ and $\Psi_{r}^{m}$ to the order of
$5n\log_{2}n$ real operations.

\section{Analysis of algorithms $\Psi_{r}^{m}$ and $\Phi_{r}^{m}$.}

In general, procedure $\Psi_{r}^{m}$ enters each end node by taking all paths
leading to this node. It turns out that the output bit error rate (BER)
significantly varies on different nodes and even on different paths leading to
the same node. Therefore our first problem is to define the most error-prone paths.

\textbf{Asymptotic analysis.} We consider AWGN channels and assume that the
all-zero codeword is transmitted as a sequence of $+1$s. Then $n$
outputs $y_{i}^{\prime}$ and $y_{i}^{\prime\prime}$ are independent
random variables $\,$(RV) with normal distribution $\mathcal{N}(1,\sigma
^{2}).$ It can be readily seen that the posterior probabilities $p_{i}$ (that
is $p_{i}^{\prime}$ or $p_{i}^{\prime\prime})$ become independent RV with
non-Gaussian distribution
\begin{equation}
p_{i}=\frac{1}{2}(1+\varepsilon_{i}),\quad\text{where }\varepsilon_{i}%
=\tanh\frac{y_{i}}{\sigma^{2}},\quad\tanh(x)\triangleq\frac{e^{x}-e^{-x}%
}{e^{x}+e^{-x}}. \label{f_calc_p_in_eps}%
\end{equation}
In the next step, we obtain the RV $p_{i}^{\mathbf{v}}$ and $p_{i}%
^{\mathbf{u}}.$ Here we rewrite equations (\ref{1}) and (\ref{3}) as follows.

\begin{lemma}
The values $p_{i}^{\mathbf{v}}$ and $p_{i}^{\mathbf{u}}$ can be calculated as
\begin{equation}
p_{i}^{\mathbf{v}}=\frac{1}{2}(1+\varepsilon_{i}^{\mathbf{v}}),\quad
\varepsilon_{i}^{\mathbf{v}}=\varepsilon_{i}^{\prime}\varepsilon_{i}%
^{\prime\prime}, \label{f_recalc_eps_v}%
\end{equation}%
\begin{equation}
p_{i}^{\mathbf{u}}=\frac{1}{2}(1+\varepsilon_{i}^{\mathbf{u}}),\quad
\varepsilon_{i}^{\mathbf{u}}=\tanh\frac{y_{i}^{\prime}+(-1)^{\widehat{v}_{i}%
}y_{i}^{\prime\prime}}{\sigma^{2}}. \label{f_recalc_eps_u}%
\end{equation}
\end{lemma}

The \emph{product }RV $\varepsilon_{i}^{\mathbf{v}}$ defined in
(\ref{f_recalc_eps_v}) has a smaller expected value relative to the original
estimates $\varepsilon_{i}^{\prime}$ and $\varepsilon_{i}^{\prime\prime},$
since $0\leq\varepsilon_{i}^{\prime},\varepsilon_{i}^{\prime\prime}\leq1$.
Therefore the mean value of $p_{i}^{\mathbf{v}}$ converges to 0.5 in the
subsequent left-hand steps. This makes decoding $\Psi_{\mathbf{v}}$ less
reliable. On the positive side, we note that each step gives us a better
protected code that has twice the relative distance of the former one.
Therefore we subsequently degrade the channel while entering the new codes
with higher correcting capabilities.

If $\widehat{v}_{i}$ is correct (i.e. $\widehat{v}_{i}=0$) we have
$\varepsilon_{i}^{\mathbf{u}}=\tanh\{(y_{i}^{\prime}+y_{i}^{\prime\prime
})/\sigma^{2}\}.$ So the second RV $\varepsilon_{i}^{\mathbf{u}}$ has a
greater expected value. Consequently, the mean probabilities $p_{i}%
^{\mathbf{u}}$ increase as we move to the right. Note, however, that each new
code has half the relative distance of its parent code. In other words, we
subsequently improve the channel while entering the new codes with weaker
correcting capabilities.

Now we proceed with an asymptotic analysis. We first consider RM codes with
$m\rightarrow\infty$ and fixed order $r$. These codes have rates
$R\rightarrow0$. Therefore we have to consider the case $\sigma^{2}%
\rightarrow\infty$ to obtain any \textit{fixed} signal-to-noise ratio
$1/R\sigma^{2}$ as $m\rightarrow\infty.$ We will use the following lemma
proven in \cite{dum3}.

\begin{lemma}
For large noise power $\sigma^{2}\rightarrow\infty$ the first two
moments $E\varepsilon$ and $E\varepsilon^{2}$ of the random variable
$\varepsilon=\tanh(y/\sigma^{2})$ satisfy the relation
\begin{equation}
E\varepsilon\sim E\varepsilon^{2}\sim\sigma^{-2}. \label{f_tanh_moments}%
\end{equation}
\end{lemma}

In general, we wish to use the original RV $y_{i},$ $\varepsilon_{i}^{\prime
},$ and $\varepsilon_{i}^{\prime\prime}$ and recalculate their probability
density functions (pdf), using (\ref{f_recalc_eps_v}) and
(\ref{f_recalc_eps_u}) to find the pdf of the new RV variables
$\varepsilon_{i}^{\mathbf{v}}$ and $\varepsilon_{i}^{\mathbf{u}}$. However,
the latter formulas make these recalculations very involved. Therefore we
consider a simplified version of our algorithm $\Psi_{\text{rec}}.$ Namely,
given a channel symbol $y$ with posterior probability $p=(1+\varepsilon)/2,$
we define the \textit{likelihood }of $0$
\[
\rho(\varepsilon)\overset{\text{def}}{=}\frac{2y}{\sigma^{2}}=\ln
\frac{1+\varepsilon}{1-\varepsilon}.
\]
Note that the likelihoods form independent Gaussian RV$.$ It can be easily
seen that the new RV $\varepsilon_{i}^{\mathbf{u}}$ obtained in
(\ref{f_recalc_eps_u}) gives the likelihood
\begin{equation}
\rho(\varepsilon_{i}^{\mathbf{u}})=\rho(\varepsilon_{i}^{\prime}%
)+\rho(\varepsilon_{i}^{\prime\prime})\label{rho1}%
\end{equation}
for any noise power $\sigma^{2}.$ For the RV $\varepsilon_{i}^{\mathbf{v}},$
the corresponding recalculation results in a longer formula
\begin{equation}
\rho(\varepsilon_{i}^{\mathbf{v}})=\ln\frac{1+\exp\rho(\varepsilon_{i}%
^{\prime})\exp\rho(\varepsilon_{i}^{\prime\prime})}{\exp\rho(\varepsilon
_{i}^{\prime})+\exp\rho(\varepsilon_{i}^{\prime\prime})}.\label{rho21}%
\end{equation}
Given the asymptotic case $\sigma^{2}\rightarrow\infty,$ note that the RV
$\rho(\varepsilon_{i})$ takes small values with high probability. Therefore we
replace the latter formula by its approximation valid for small
$\rho(\varepsilon_{i}^{\prime})$ and $\rho(\varepsilon_{i}^{\prime\prime})$:
\begin{equation}
\rho(\varepsilon_{i}^{\mathbf{v}})\sim\rho(\varepsilon_{i}^{\prime}%
)\rho(\varepsilon_{i}^{\prime\prime}).\label{rho2}%
\end{equation}
It can be proven that the output bit error rate obtained on any end node
$\left\{
\genfrac{}{}{0pt}{}{g}{0}%
\right\}  $ with large $g$ can be calculated using only the first two moments
$E\rho$ and $E\rho^{2}$ of the RV $\rho=\rho(\varepsilon_{i}^{\mathbf{v}})$
obtained on this node. It can also be proven that the original formula
(\ref{rho21}) and its approximation (\ref{rho2}) give the same moment $E\rho$
as $\sigma^{2}\rightarrow\infty.$ Also, the two formulas give the same
asymptotic moments  $E\rho^{2}.$ This justifies using the above approximation
in the asymptotic case$.$

Now we consider a simplified algorithm $\Psi_{\text{rec}},$ with
recalculations (\ref{f_recalc_eps_v}) and (\ref{f_recalc_eps_u}) replaced by
(\ref{rho1}) and (\ref{rho2}). Using this simplified version $\Psi
_{\text{rec}},$ we can arrive at the following conclusions \cite{dum4}.

\begin{theo}
For $\sigma^{2}\rightarrow\infty$, replacing code $\left\{
\genfrac{}{}{0pt}{}{m}{r}%
\right\}  $ by $\left\{
\genfrac{}{}{0pt}{}{m-1}{r-1}%
\right\}  $ in the algorithm $\Psi_{\mathbf{v}}$ is equivalent to increasing
the original noise power $\sigma^{2}$ to $\sigma^{4}$. Replacing code $\{%
\genfrac{}{}{0pt}{}{m}{r}%
\}$ by $\left\{
\genfrac{}{}{0pt}{}{m-1}{r}%
\right\}  $ in the algorithm $\Psi_{\mathbf{u}}$ reduces the original noise
power $\sigma^{2}$ to $\sigma^{2}/2.$
\end{theo}

\begin{figure}[htb]
\begin{center}
\includegraphics[width=\linewidth]{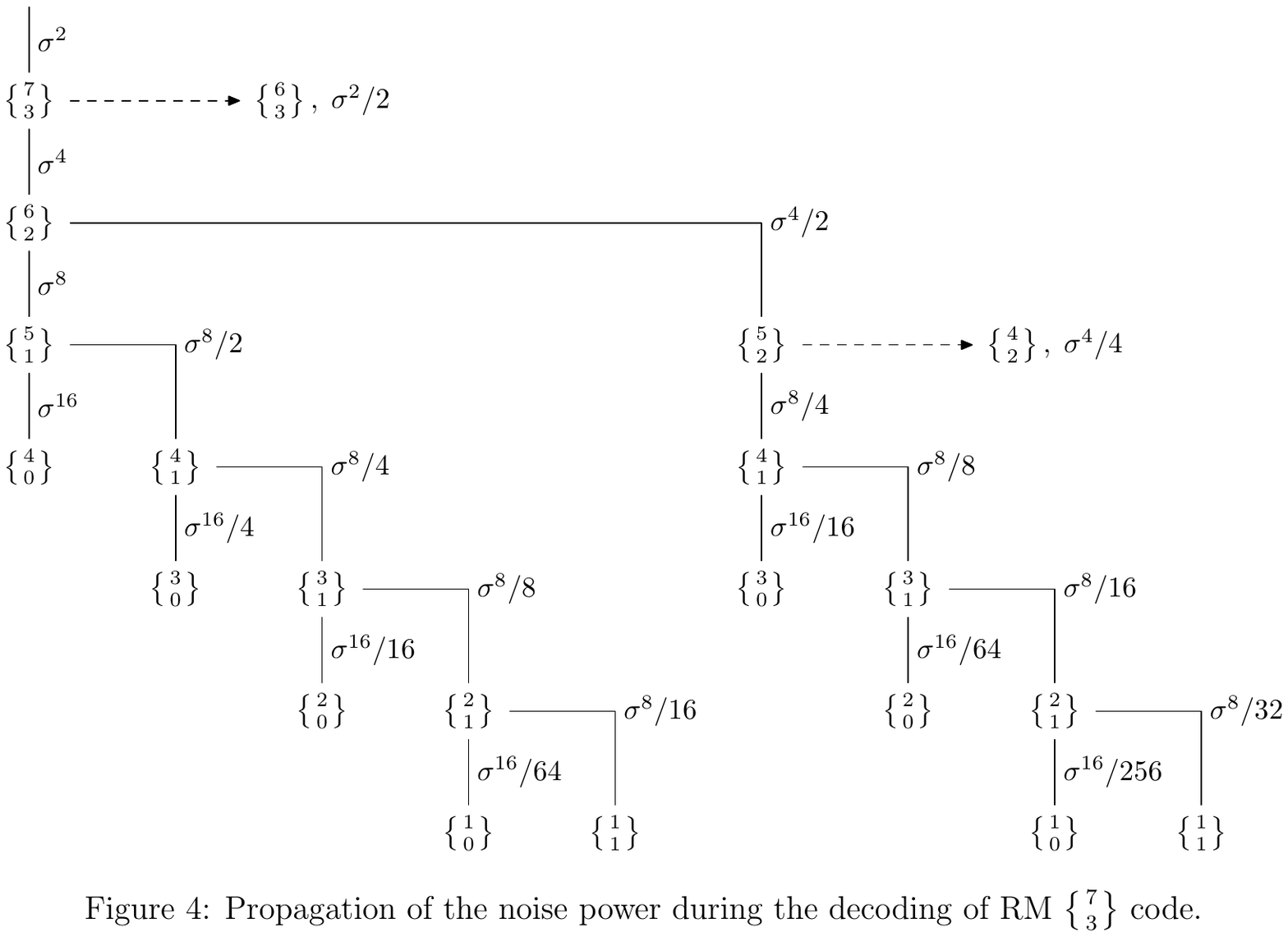}
\end{center}
\label{ds_fig3}
\end{figure}

Therefore in asymptotic setting our recursive procedure can be considered as a
``propagation of the noise power''. This propagation undergoes two different
types of changes while we move from the parent node to the two descendant
nodes. This propagation is also illustrated on Fig.~\ref{ds_fig3} as an example for the
$\left\{\genfrac{}{}{0pt}{}{7}{3}\right\}$
code (note, however, that procedure becomes exact only for very
long codes).
\setcounter{figure}{4}
Now we can find asymptotic error rate for each bit (path) $a_{i}$. Note that
algorithm $\Psi_{r}^{m}$ has the highest noise power $\sigma^{2^{r+1}}$ when
it arrives at the leftmost (repetition) code $\left\{
\genfrac{}{}{0pt}{}{m-r}{0}%
\right\}  .$ Also, a repetition code of any length $n$ used on the AWGN
channel $\mathcal{N}(0,\sigma^{2})$ has an output error probability of ML
decoding
\begin{equation}
P=Q\left(  \sqrt{n}/\sigma\right)  ,\quad\text{where }Q(x)\triangleq\int
_{x}^{\infty}e^{-u^{2}/2}\,du\,/\sqrt{2\pi}.\label{f_perr_rep}%
\end{equation}
Note that for a general binary memoryless channel with noise variance
$\sigma^{2},$ the same estimate can also be used on any node $\left\{
\genfrac{}{}{0pt}{}{g}{0}%
\right\}  $ given the following two conditions:
\begin{equation}
g\rightarrow\infty,\quad\sigma/2^{g/3}\rightarrow\infty.\label{sums}%
\end{equation}
Both conditions directly follow from the estimates obtained in \cite{fel}
(p.~549) for the large deviations of the sums of independent random variables with
common distribution.

Estimate (\ref{f_perr_rep}) shows that the algorithm $\Psi_{r}^{m}$ gives
the highest error rate
\[
\mathcal{P}_{0}=Q\left(  \sqrt{2^{m-r}}/\sigma^{2^{r}}\right)
\]
on the leftmost node $\left\{
\genfrac{}{}{0pt}{}{m-r}{0}%
\right\}  $. The second highest error rate $\mathcal{P}_{1}=Q\left(
\sqrt{2^{m-r+1}}/\sigma^{2^{r}}\right)  $ is obtained on the next node
$\left\{
\genfrac{}{}{0pt}{}{m-r-1}{0}%
\right\}  .$ Note that these probabilities rapidly decline. In particular,
$\mathcal{P}_{1}\sim\mathcal{P}_{0}^{2}$ when $\mathcal{P}_{0}$ is small. In
fact, for most noise powers $\sigma^{2},$ the first BER $\mathcal{P}_{0}$
exceeds all subsequent bit error rates so considerably that it practically
defines the overall word ER (WER). By contrast, the lowest BER is obtained on
the rightmost node $\left\{
\genfrac{}{}{0pt}{}{m-r}{m-r}%
\right\}  .$ Thus, we arrive at the following conclusions:

$\bullet$ The left-hand movement from a code $\left\{
\genfrac{}{}{0pt}{}{j}{i}%
\right\}  $ to the next code $\left\{
\genfrac{}{}{0pt}{}{j-1}{i-1}%
\right\}  $ increases the output BER. In this case, doubling the relative code
distance $d/n$ does not compensate for a stronger noise obtained on the code
$\left\{
\genfrac{}{}{0pt}{}{j-1}{i-1}%
\right\}  .$

$\bullet$ Moving to the right from a code $\left\{
\genfrac{}{}{0pt}{}{j}{i}%
\right\}  $ to the next code $\left\{
\genfrac{}{}{0pt}{}{j-1}{i}%
\right\}  $ allows us to reduce the BER of the algorithm relative to the
parent code $\left\{
\genfrac{}{}{0pt}{}{j}{i}%
\right\}  $. As a result, the lowest BER is obtained on the rightmost node
$\left\{
\genfrac{}{}{0pt}{}{r}{r}%
\right\}  .$\medskip

In a more general setting, we can estimate asymptotic error rates for any
information bit $a_{j}$. For $c=0,1$ and $x>0,$ we use notation
$x \diamond c \overset{\text{def}}{=} 2^{1-c}x^{1+c}.$
Given a symbol $a_{j}$ arriving at the node
$\left\{\genfrac{}{}{0pt}{}{g}{0}\right\}$
we can consider the corresponding subpath
$\overline{j}_{g} = (j_{1},\ldots,j_{m-g})$.
Then we define the product
$x\diamond\overline{j}_{g}=(\cdots(x\diamond j_{1})\diamond j_{2}\cdots) \diamond j_{m-g}$
and arrive at the following statement.

\begin{theo}
Consider RM codes with $m\rightarrow\infty$ and fixed order $r$. For the
information bit $a_{j}$ associated with a node
$\left\{\genfrac{}{}{0pt}{}{g}{0}\right\}$,
algorithm $\Psi_{r}^{m}$ has bit error rate
\begin{equation}
\mathcal{P(}a_{j}) \sim Q\left(  \sigma\diamond\overline{j}_{g}\right)  ,\quad
g\rightarrow\infty. \label{f_dec_RM_r0_p1}%
\end{equation}
\end{theo}

Similar results hold for the algorithm $\Phi_{r}^{m}$, which stops at the
nodes $\left\{\genfrac{}{}{0pt}{}{g}{1}\right\}$.
This node is associated with a subblock of $g+1$ information
bits. In this case the corresponding subindex $\overline{j}_{g}^{1}$ has
weight $r-1$ or less. Therefore algorithm $\Phi_{r}^{m}$ reduces the highest
noise power $\sigma^{2^{r+1}}$ to $\sigma^{2^{r}}$ and gives substantial
improvement over $\Psi_{r}^{m}.$ More generally, we obtain the following statement.

\begin{theo}
Consider RM codes with $m\rightarrow\infty$ and fixed order $r$. For the subset
of $g+1$ information bits $\mathcal{\{}a_{j}\}$ associated with a node
$\left\{\genfrac{}{}{0pt}{}{g}{1}\right\}$,
algorithm $\Psi_{r}^{m}$ has bit error rate
\[
\mathcal{P(\{}a_{j}\}) \lesssim
2^{g} Q\left(\sigma\diamond\overline{j}_{g}^{1}\right),
\quad g\rightarrow\infty.
\]
\end{theo}

In particular, the highest BER $\mathcal{P}_{0}$ obtained at the node
$\left\{\genfrac{}{}{0pt}{}{m-r}{0}\right\}$
by $\Psi_{r}^{m}$ is now being replaced by
\[
\mathcal{P}_{0}^{\prime} \lesssim 2^{m-r+1} Q\left(  \sqrt{2^{m-r}}/\sigma^{2^{r-1}%
}\right)
\]
obtained at the node $\left\{\genfrac{}{}{0pt}{}{m-r+1}{1}\right\}$.
As the block length grows, decoding $\Phi_{r}^{m}$
increasingly outperforms both the majority algorithm and recursive
techniques of \cite{lit}, \cite{kab}. Further, this analysis can be extended
to codes of fixed rate $R.$ In particular, the following statement holds for
hard-decision decoding.

\begin{theo}
For RM codes with $m\rightarrow\infty$ and fixed rate $R$, algorithm
$\Phi_{r}^{m}$ has error-correcting threshold $(d\ln d)/2$ thus:

$\bullet$ increasing $\ln d$ times the threshold of bounded-distance decoding;

$\bullet$ doubling the threshold $(d\ln d)/4$ of majority decoding.
\end{theo}

\section{Improvements}

\subparagraph{1. Subcodes of RM codes}

To improve output error rate, we \textit{set
the leftmost information bits as zeros}. In this way, we arrive at the
subcodes of the original code $\left\{\genfrac{}{}{0pt}{}{m}{r}\right\}$
that are obtained by eliminating a few least protected information
bits. This expurgation starts with the node
$\left\{\genfrac{}{}{0pt}{}{m-r}{0}\right\}$
in procedure $\Psi_{r}^{m}$, and with the node
$\left\{\genfrac{}{}{0pt}{}{m-r+1}{1}\right\}$
in $\Phi_{r}^{m}$. It can be shown that after eliminating only
one bit, algorithm $\Psi_{r}^{m}$ gives the same BER on the channel whose
noise power $\sigma^{2}$ is increased $2^{1/2^{r}}$ times. For the algorithm
$\Phi_{r}^{m},$ the sustainable noise power is increased $2^{1/2^{r-1}}$
times. For long codes of small order $r=2,3$ this amounts to a gain of 1.5
dB and 0.75 dB, respectively.

\subparagraph{2. List decoding}

Decoding performance is further improved by
choosing $L$ best candidates after each decoding step. This \textit{list
decoding} $\Psi_{r}^{m}(L)$ starts at the leftmost code
$\left\{\genfrac{}{}{0pt}{}{m-r}{0}\right\}$.
Here we define posterior probabilities $p(\mathbf{v} \mid \mathbf{y})$
of both codewords $\mathbf{v}^{\prime} = 0$ and $\mathbf{v}^{\prime\prime} = 1$.
These codewords are represented as two
initial edges with the corresponding cost functions
$\log p(\mathbf{v} \mid \mathbf{y})$.
Then we decode the next code
$\left\{\genfrac{}{}{0pt}{}{m-r-1}{0}\right\}$.
Note that codewords $\mathbf{v}$\textbf{$^{\prime}$} and
$\mathbf{v}^{\prime\prime}$ give different probability distributions on this
node. Therefore our new decoding is performed 2 times, separately for
\textbf{$\mathbf{v}^{\prime}$} and $\mathbf{v}^{\prime\prime}\mathbf{.}$
The result is a full tree of depth 2 that has 4 new edges. On further
steps, we keep doubling the number of paths until $2L$ paths are formed. Then
we choose $L$ paths with maximum cost functions and proceed further. In the
end, the most probable path (that has maximum cost function) is chosen among
$L$ paths survived at the rightmost node. Simulation results and analytic
estimates give very substantial improvements when both techniques -- using the
subcodes and short decoding lists -- are combined. These results are presented
below in Figures \ref{ds_fig4} to \ref{ds_fig7}.

\subparagraph{3. New permutation techniques}

Finally, the third improvement
utilizes the rich \textit{symmetry group} $GA(m)$ of RM codes that includes
$2^{O(m^{2})}$ permutations. First, note that even for large $L$ algorithm
$\Psi_{r}^{m}(L)$ is likely to fail if error positions substantially
disagree on the two halves of the original block. By using a symmetry group,
we try to find the permutations that match unknown erroneous positions in the
two permuted halves. If successful, procedure $\Psi_{\mathbf{v}}$ will
eliminate most errors on the permuted block $\mathbf{v}$ of length $n/2.$ This
process can be advanced in the next step, by finding another permutation that
again gives a good match on erroneous positions left on the new halves of
length $n/4.$

In particular, we use the following sets of permutations. Represent any
position $i=1,\ldots,2^{m}$ in the binary form $i=(i_{1},\ldots,i_{m}).$ We take any
permutation $\pi(1),\ldots,\pi(m)$ and consider the subgroup $S$ $\subset
GA(m)$ that includes $m!$ permutations $(i_{1},\ldots,i_{m})\mapsto(i_{\pi
(1)},\ldots,i_{\pi(m)}).$ Note that using subgroup $S$ also changes the
``folding'' order used in algorithm $\Psi_{\mathbf{v}}$ (say, we fold adjacent
quarters instead of halves when we permute $i_{1}$ and $i_{2}).$ We can also
consider $\left(\genfrac{}{}{0pt}{}{m}{r}\right)$
permutations taking exactly one permutation with a given subset
$\pi^{-1}(1),\ldots,\pi^{-1}(r)$ of the first $r$ ``folding'' indices.
Permutations from this subset $T$ change the order in which we decode
left-end nodes. Finally, consider a subgroup $U$ that includes $m$ cyclic
shifts $\pi(1),\ldots,\pi(m).$ Simulation results for the moderate lengths 256
and 512 showed that using subsets $T$ and even $U$ allows to reduce the
\textit{combined} list of $L$ best candidates by one decimal order. As a
result, we obtained nearly ML decoding on the lengths 512 while using lists of
moderate size $L.$

\section{Simulation results}

\begin{figure}[htb]
\begin{center}
\includegraphics[width=\linewidth]{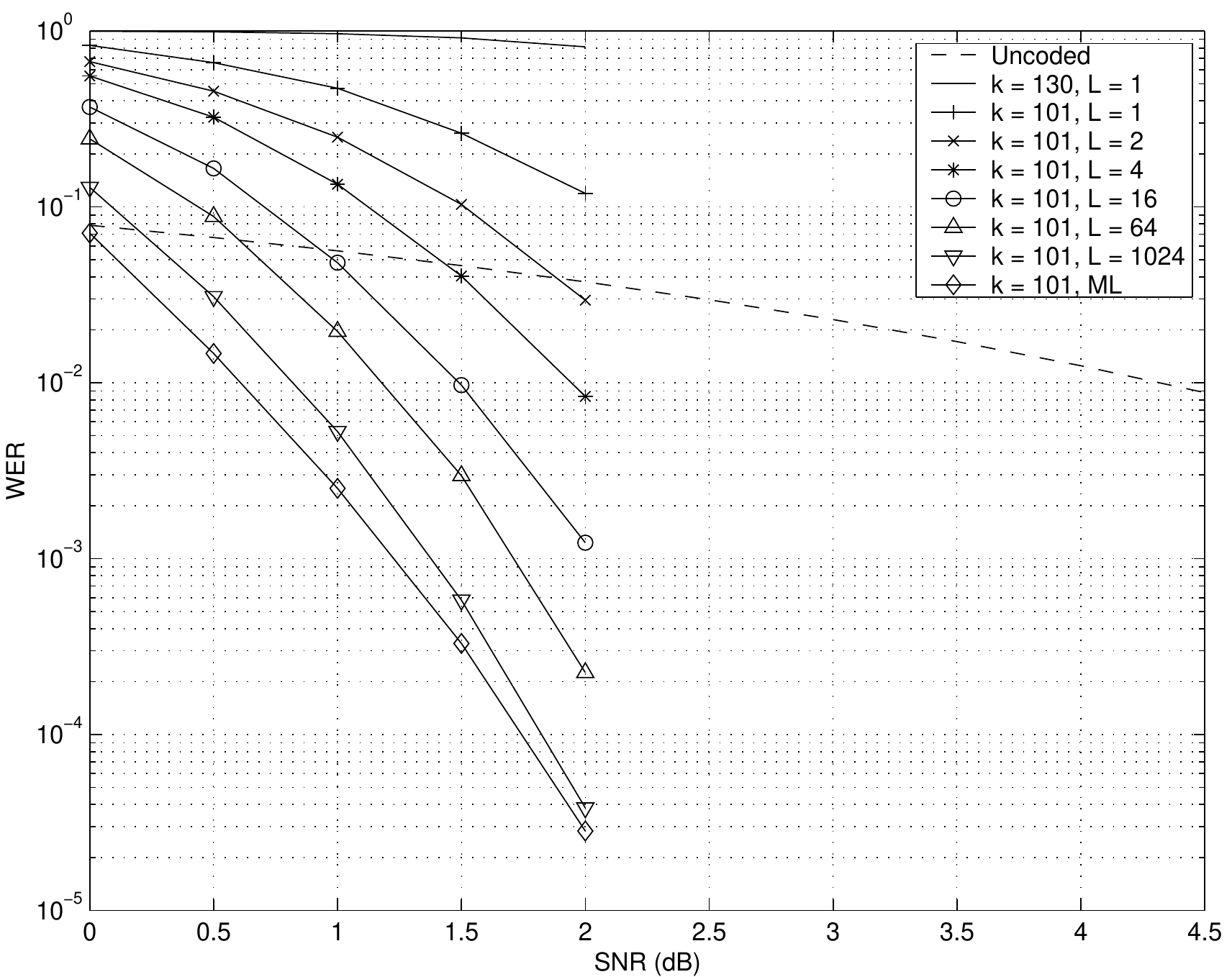}
\caption{RM $\left\{\genfrac{}{}{0pt}{}{9}{3}\right\}$ code, $n = 512$, $k = 130$,
and its subcode, $k = 101$.}
\label{ds_fig4}
\end{center}
\end{figure}

\begin{figure}[htb]
\begin{center}
\includegraphics[width=\linewidth]{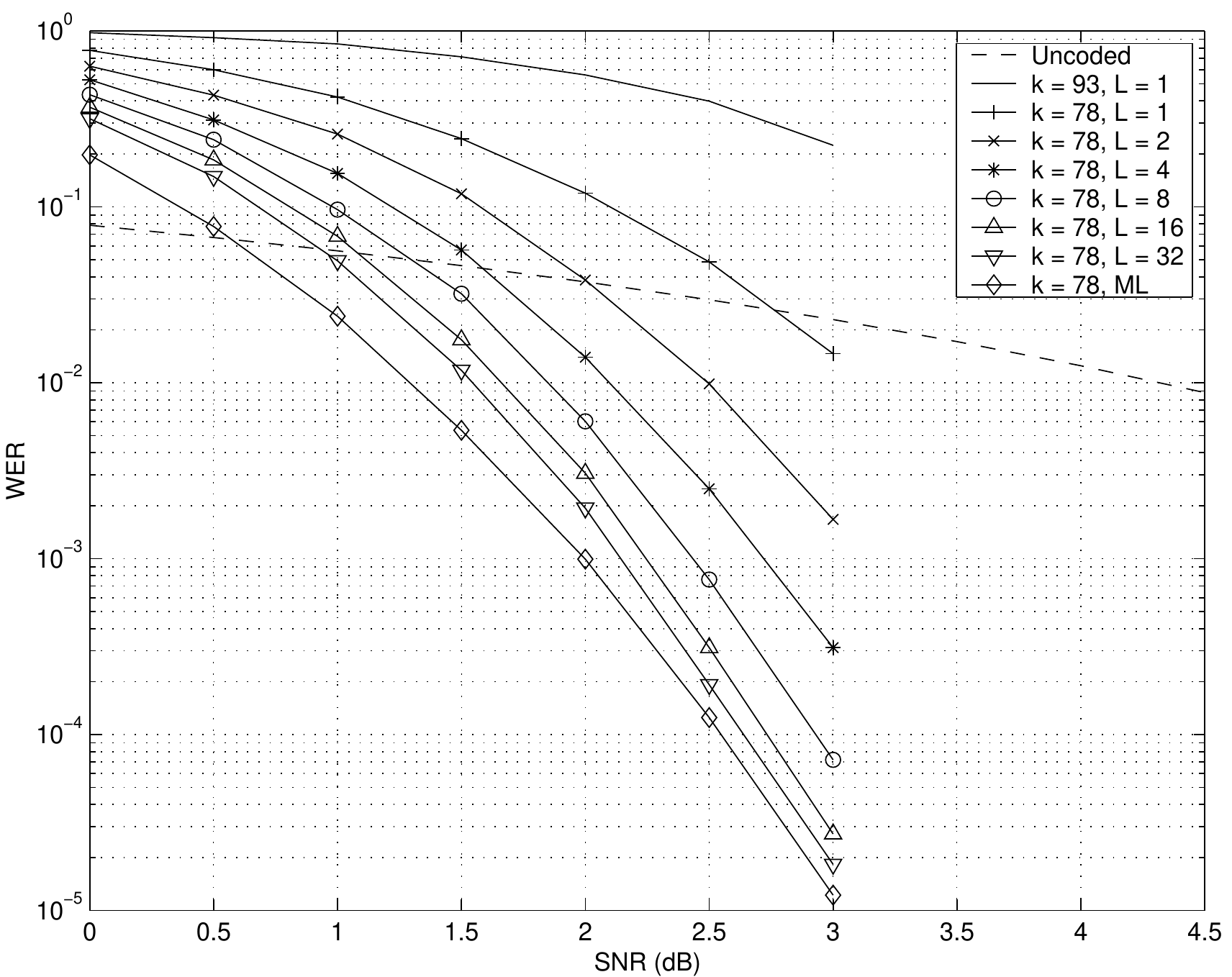}
\caption{RM $\left\{\genfrac{}{}{0pt}{}{8}{3}\right\}$ code, $n = 256$, $k = 93$,
and its subcode, $k = 78$.}
\label{ds_fig5}
\end{center}
\end{figure}

\begin{figure}[htb]
\begin{center}
\includegraphics[width=\linewidth]{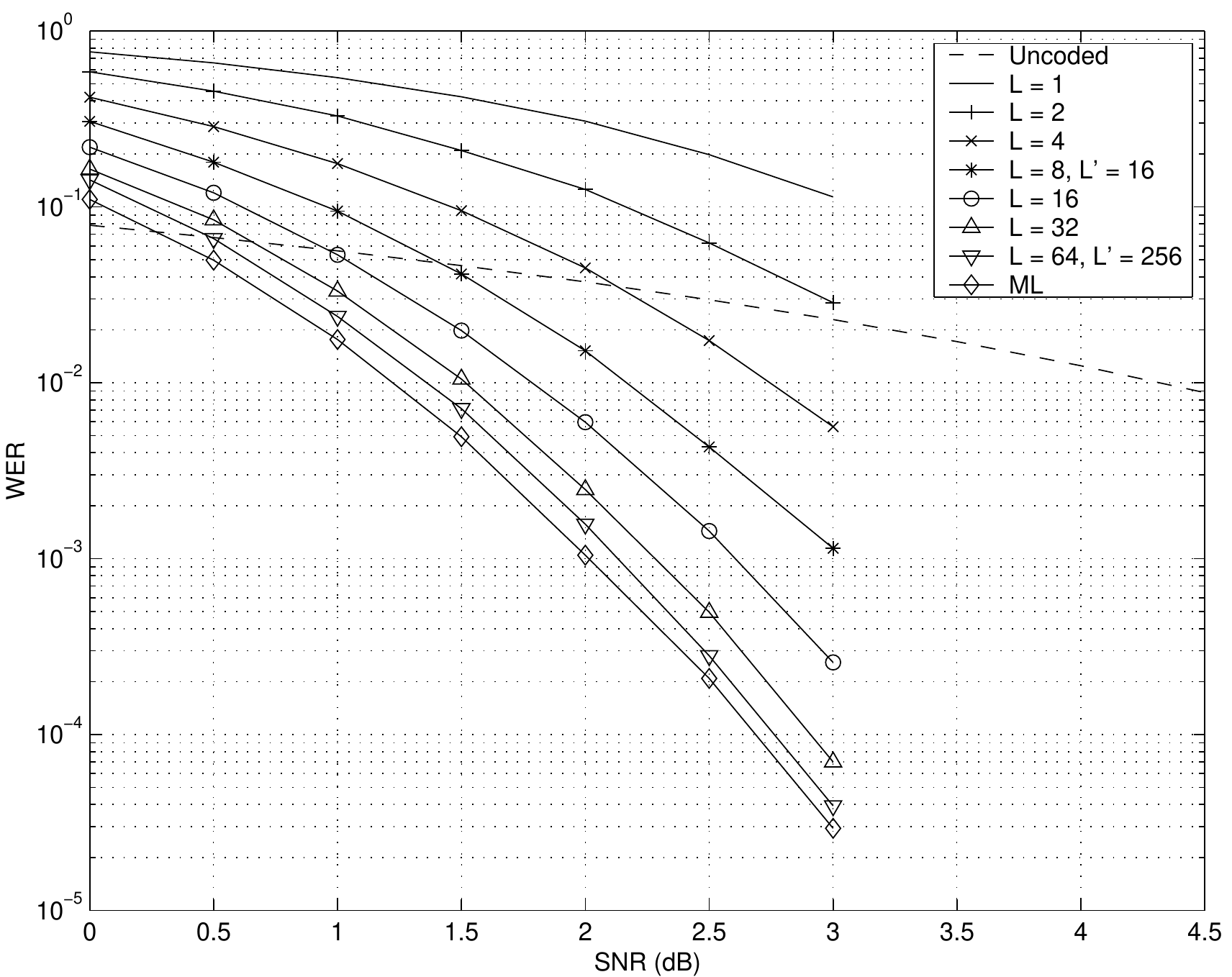}
\caption{RM $\left\{\genfrac{}{}{0pt}{}{8}{2}\right\}$ code, $n = 256$, $k = 37$.
$L'$ is the number of paths required to achieve approximately the same WER
when no permutations are used.}
\label{ds_fig6}
\end{center}
\end{figure}

\begin{figure}[htb]
\begin{center}
\includegraphics[width=\linewidth]{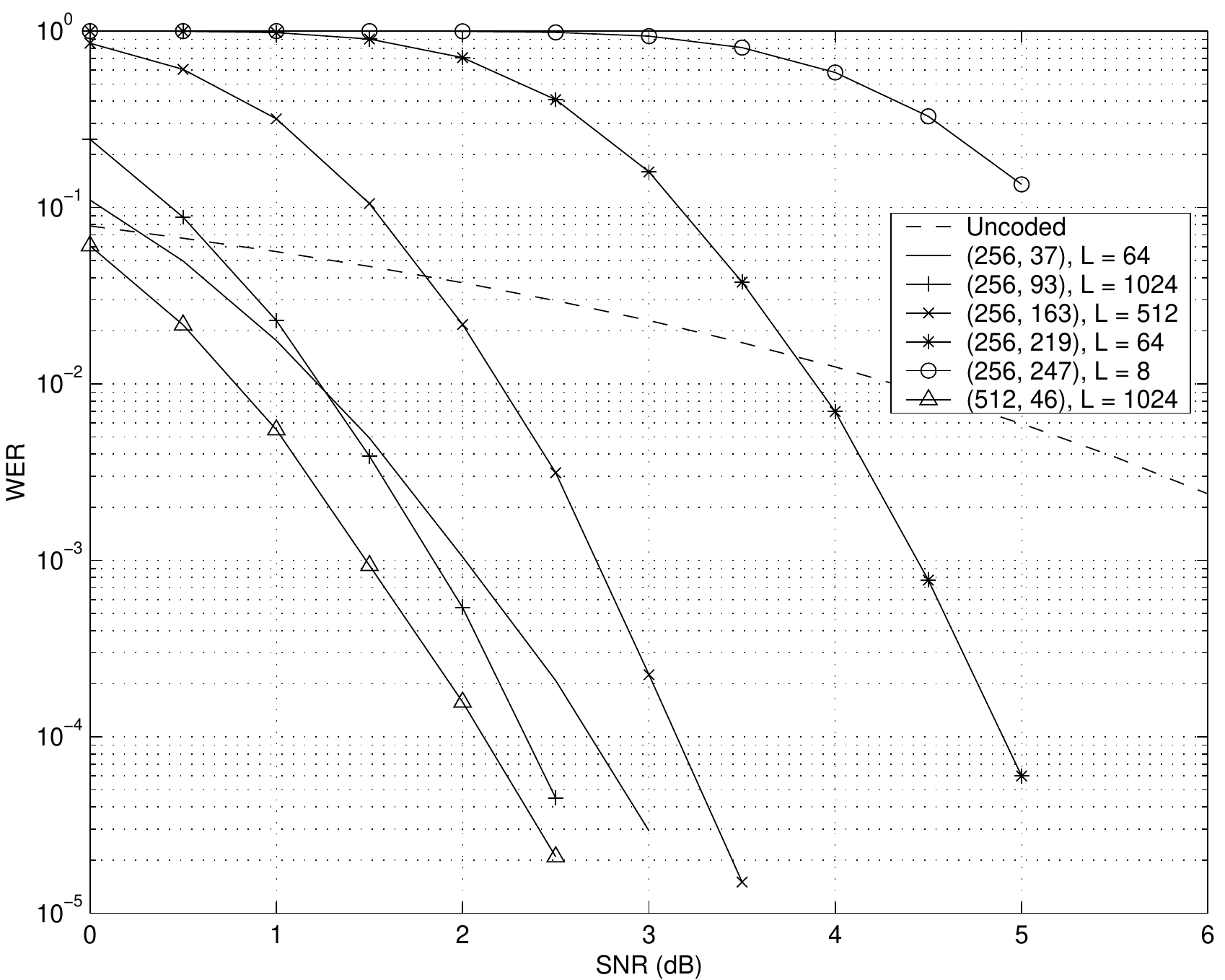}
\caption{Experimental lower bounds for RM
$\left\{\genfrac{}{}{0pt}{}{8}{2}\right\}, \ldots,
\left\{\genfrac{}{}{0pt}{}{8}{6}\right\}$, and
$\left\{\genfrac{}{}{0pt}{}{9}{2}\right\}$ codes.}
\label{ds_fig7}
\end{center}
\end{figure}

Simulation results are described in Figures \ref{ds_fig4} to \ref{ds_fig7}.
We start with Fig.~\ref{ds_fig4}
that reflects substantial improvements obtained when both techniques --
using the subcodes and short decoding lists -- were combined. The first (upper)
curve with $L=1$ shows the performance of the algorithm $\Psi_{r}^{m}$
applied to the $\left\{\genfrac{}{}{0pt}{}{9}{3}\right\}$
code with $n=512$ and $k=130.$ This algorithm can be considered as
a refined version of the former recursive techniques from \cite{lit},
\cite{kab}, and \cite{bos}. Namely, $\Psi_{r}^{m}$ uses exact probability
recalculations presented in formulas (\ref{f_recalc_eps_v}) and
(\ref{f_recalc_eps_u}) instead of various metrics used in these papers.

The second curve with $L=1$ shows the performance of the algorithm $\Psi
_{r}^{m}$ applied to the $(512,101)$-subcode of the original
$\left\{\genfrac{}{}{0pt}{}{9}{3}\right\}$ code.
This subcode is obtained by removing 29 leftmost information
bits with the highest BER. We see that the subcode gives substantial
improvement in the output BER despite having a smaller code rate. All other
curves on Fig.~\ref{ds_fig4} correspond to the same subcode
but use the bigger lists.
We see from Fig.~\ref{ds_fig4} that algorithm $\Psi_{r}^{m}$ is further improved
by 3.5 to 5 dB at BER $10^{-4},$ by using the algorithm $\Psi_{r}^{m}(L)$ with
moderate number $L$.

For large $L,$ simulation results (exhibited in Fig.~\ref{ds_fig4} and \ref{ds_fig5})
also showed
that most incorrectly decoded codewords are still more probable than the
transmitted vector. This fact shows that our word ER (WER) is very close
to that of ML decoding. In turn, this gives a new (experimental) bound on the
WER of ML decoding.

It is also interesting that subcodes usually achieve near-ML decoding using
much smaller lists relative to the original RM codes. In particular, Fig.~\ref{ds_fig5}
presents simulation results for a (256,78)-subcode of the (256,93)-code
$\left\{\genfrac{}{}{0pt}{}{8}{3}\right\}$.
This subcode approaches near-ML decoding using \textit{only 32
intermediate paths, }while the original
$\left\{\genfrac{}{}{0pt}{}{8}{3}\right\}$
requires about 512 paths (using permutation techniques) or even
4096 paths (without permutations). Note that even one of the most efficient
algorithms developed in \cite{hart2} uses about $10^{5}$ paths for BCH codes
of length 256.

Fig.~\ref{ds_fig6}
presents simulation results when permutation techniques were applied to
the $\left\{\genfrac{}{}{0pt}{}{8}{2}\right\}$
code with $n=256$ and $k=37.$ Here we compare the original
recursive algorithm $\Psi_{r}^{m}(L)$ with its refined version that uses a
small subgroup $U$ defined in the previous section. The results show that
adding a few permutations can substantially reduce the overall list size
(taken over all permutations). For this specific code, the refined version
reduces approximately 4 times the number of trials used in $\Psi_{r}^{m}(L)$
to obtain near-ML decoding. Similar results show that for codes of length
$512$ the complexity of near-ML decoding is reduced tenfold.

Finally, in Fig.~\ref{ds_fig7},
we summarize the results for all nontrivial RM codes of
length 256 and for the code
$\left\{\genfrac{}{}{0pt}{}{9}{2}\right\}$
of length 512. This figure presents almost exact experimental
bounds on the error probability of ML decoding, along with the minimum lists
$L$ that were used to meet ML-decoding performance. Here we also use
permutation techniques to reduce this size $L.$ An interesting open problem is
to provide a theoretical explanation as to why permutation decoding allows to
substantially reduce the overall size $L$ of the lists over the basic
recursive algorithms $\Psi_{r}^{m}$ and $\Phi_{r}^{m}$.

\section{Conclusion}

Our main conclusion is that recursive decoding of RM codes combines good
performance and low complexity on \textit{moderate }blocklengths up to 512. In
turn, this allows us to partially fill the gap left by optimum maximum
likelihood (ML) decoding and suboptimal iterative decoding. Note that the
former has unfeasible complexity for nontrivial codes even on relatively short
blocks of hundreds bits, while the latter becomes very efficient for turbo
codes and low parity-check codes only on the blocks of thousands bits. An
important open problem is whether recursive techniques can enable fast near-ML
decoding for the lengths of 1024 and 2048. A positive solution to this problem
would allow to completely fill the gap in the blocklengths left by the best
algorithms known to date.


\end{document}